\documentclass[10pt,journal]{IEEEtran} 

\usepackage[utf8]{inputenc}
\usepackage[T1]{fontenc}
\usepackage{amssymb,amsmath}
\usepackage{cite}
\usepackage{psfrag}
\usepackage{url}
\usepackage[absolute,overlay]{textpos}
\DeclareMathOperator{\Tr}{Tr}
\usepackage{pgf}
\usepackage[export]{adjustbox}
\usepackage{bm}
\usepackage{bbm}
\usepackage{booktabs}
\usepackage{url}
\usepackage{upgreek}
\usepackage{verbatimbox}
\usepackage{algorithm}
\usepackage{algorithmic}
\usepackage{enumitem}

\newcommand{\appropto}{\mathrel{\vcenter{
			\offinterlineskip\halign{\hfil$##$\cr
				\propto\cr\noalign{\kern2pt}\sim\cr\noalign{\kern-2pt}}}}}

\usepackage{amsthm}

\begin{document}
	
	\title{CSI-free Rotary Antenna Beamforming for\\ Massive RF Wireless Energy Transfer}	
	
	\author{ Onel L. A. López,~\IEEEmembership{Member,~IEEE,}
	Hirley Alves,~\IEEEmembership{Member,~IEEE,}
	Samuel Montejo-S\'anchez,~\IEEEmembership{Member,~IEEE,}
	Richard D. Souza,~\IEEEmembership{Senior Member,~IEEE}
		and Matti Latva-aho,~\IEEEmembership{Senior Member,~IEEE}
		\thanks{Onel López, Hirley Alves, and Matti Latva-aho are  with  the Centre for Wireless Communications University of Oulu, Finland, e-mails: \{Onel.AlcarazLopez, Hirley.Alves, Matti.Latva-aho\}@oulu.fi.}
		\thanks{Samuel Montejo-S\'anchez is with Programa Institucional de Fomento a la I+D+i, Universidad Tecnol\'ogica Metropolitana, Santiago, Chile, e-mail: smontejo@utem.cl.}
		\thanks{Richard D. Souza is with Federal University of Santa Catarina (UFSC), 	Florian\'opolis, Brazil, richard.demo@ufsc.br.}
		\thanks{This work has been partly funded by the European Commission through the H2020 project Hexa-X (Grant Agreement no. 101015956). This research has been financially supported by Academy of Finland through 6Genesis Flagship (Grant no. 318927) and EE-IoT (no. 319008), by RNP/MCTIC (Grant 01245.010604/2020-14) 6G Mobile Communications Systems, Brazil, and by ANID FONDECYT Iniciación no. 11200659, Chile.}
	} 
	
	\maketitle
	
	\begin{abstract}
    Radio frequency (RF) wireless energy transfer (WET) is a key technology that may allow  seamlessly powering future massive low-energy Internet of Things (IoT) networks. To enable efficient massive WET, channel state information (CSI)-limited/free multi-antenna transmit schemes have been recently proposed in the literature. The idea is to reduce/null the energy costs to be paid by energy harvesting (EH) IoT nodes from participating in large-scale time/power-consuming CSI training, but still enable some transmit spatial gains. In this paper, we take another step forward by proposing a novel CSI-free rotary antenna beamforming (RAB) WET scheme that outperforms all state-of-the-art CSI-free schemes in a scenario where a power beacon (PB) equipped with a uniform linear array (ULA) powers a large set of surrounding EH IoT devices. 
    RAB uses a properly designed CSI-free beamformer combined with a continuous or periodic rotation of the ULA at the  PB to provide average EH gains that scale as $0.85\sqrt{M}$, where $M$ is the number of PB's antenna elements. Moreover, a rotation-specific power control mechanism was proposed to i) fairly optimize the WET process if devices' positioning information is available, and/or ii) to avoid hazards to human health in terms of specific absorption rate (SAR), which is an RF exposure metric that quantifies the absorbed power in a unit mass of human tissue.    
    We show that RAB performance even approaches quickly (or surpasses, for scenarios with sufficiently large number of EH devices, or when using the proposed power control) the performance of a traditional full-CSI based transmit scheme, and it is also less sensitive to SAR constraints. Finally, we discuss important practicalities related to RAB such as its robustness against non line-of-sight conditions compared to other CSI-free WET schemes, and its generalizability to scenarios where the PB uses other than a ULA topology. 
	\end{abstract}
	\section{Introduction}\label{intro}
	Radio frequency (RF) wireless energy transfer (WET) technology, hereinafter referred just as WET, is widely recognized as a green enabler of low-power Internet of Things (IoT). WET realizes battery charging without physical connections (thus, simplifying servicing and maintenance, and contributing to reduce waste processing); and favors form factor reduction and durability increase of the end devices \cite{Lopez.2021}.		
	Moreover, WET solutions are attractive for wirelessly energizing future extremely massive IoT systems with deployment densities reaching tens devices per square meter (devices/$\mathrm{m}^2$) \cite{Mahmood.2020}, for which network planning, maintenance and operation should be made robust, but also simple and flexible.
	In fact, WET commercialization is already gaining momentum with a variety of emerging enterprises with a large portfolio of technical solutions, e.g., Powercast, TransferFi and Ossia,\footnote{See \url{https://www.powercastco.com}, \url{https://www.transferfi.com} and \url{https://www.ossia.com}.} including the provision of dedicated RF transmitters for WET, i.e., power beacons (PBs), for wirelessly charging low-power IoT devices. 
	
	Supporting the powering of massive IoT deployments via WET, which is referred in the literature as massive WET \cite{Lopez.2021}, still constitutes a relatively novel and challenging problem. 
	Among the main approaches to enable massive WET are the densification and appropriate deployment of PBs \cite{Wang.2016,Dai.2018,Liang.2019,Arivudainambi.2018,Rosabal.2020} and/or increasing the number of antennas per PB to leverage high spatial gains \cite{Kashyap.2016,Wang.2019,Khan.2018}, which in turn are costly and not always possible. 	
	Specifically, authors in \cite{Wang.2016} proposed an adaptive directional WET scheme, where the PBs adapt the antenna beam direction according to the location of the EH devices, and analyzed the statistics of the receive power using stochastic geometry. Likewise, another directional charging model is developed in \cite{Dai.2018} but to optimize a network utility function, i.e., the network average harvested energy.  
	Meanwhile, a non-uniform PBs deployment strategy is designed in \cite{Liang.2019} such that a base station (BS) and the multiple PBs can cooperate to power the entire set of energy harvesting (EH) devices.
	They derived the network total power consumption to satisfy a certain energy coverage probability requirement, and showed the important energy savings coming from the cooperation of the BS in the WET process. In \cite{Arivudainambi.2018}, a Daubechies wavelet algorithm is proposed to identify the optimal PBs positions aiming to improve the  area charging coverage to a massive number of sensors. Meanwhile, authors in \cite{Rosabal.2020} optimized both, the number of PBs and their positions, to support worst-case charging coverage without devices positioning information at the network side. They showed that the number of deployed PBs play a much more relevant role for performance improvement than the number of antennas per-PB. On the other hand, the potential benefits of using massive antenna arrays for WET are discussed in \cite{Kashyap.2016}, while an energy efficiency optimization for massive multiple-input
	multiple-output (MIMO) systems considering the hardware impairments at the EH sensors is carried out in \cite{Wang.2019}. 
	The   overall power transfer efficiency of a massive MIMO system is investigated in \cite{Khan.2018} as a function of the number of transmit antennas and number of EH devices.

	In general, PBs may be equipped with directional antennas, e.g., \cite{Wang.2016,Dai.2018}, thus, the coverage regions must be carefully planned; or omnidirectional antennas, e.g., \cite{Liang.2019,Arivudainambi.2018,Rosabal.2020,Kashyap.2016,Wang.2019,Khan.2018}, thus, multiple antennas are required to beamform the energy into the desired spatial directions \cite{Alsaba.2018}. 	
	In the latter scenario, which is the most common in the literature due to its flexibility, each PB traditionally uses some form of channel state information (CSI) of the WET channels to maximize the energy transfer. However, CSI acquisition is already challenging in WET-enabled small-scale networks, let alone in massive IoT deployments \cite{Lopez.2021}. The reasons are that accurate CSI acquisition requires significant amounts of time and energy from the EH devices, which may erase
	(or even reverse) the gains from energy beamforming \cite{Zeng.2015,Flordelis.2018,Lopez.2020}. This is why authors in \cite{Zeng.2015} proposed a framework for optimizing the net harvested energy, which is the average  energy harvested by a multi-antenna device offset by that used for channel training.\footnote{In general, time-division duplexing systems exploiting channel reciprocity are often more robust again CSI training related costs than frequency-division duplexing systems \cite{Flordelis.2018}, specially in WET-enabled setups \cite{Lopez.2020}.}  	
	Moreover, not only CSI training and corresponding energy expenditure scales up exponentially in massive IoT deployments, thus, becoming extremely costly and challenging \cite{Lopez.2020}, but also the (possible) gains from energy beamforming decrease quickly as the number of served devices increases. 
	
	To overcome the above limitations imposed by CSI acquisition in (possibly massive) WET setups, novel efficient CSI-limited/free WET schemes have been recently proposed in \cite{LopezMonteiro.2021,Clerckx.2018,LopezAlves.2019,LopezMontejo.2020}. Specifically, a low-complexity WET beamformer exploiting only the average statistics of the channel was proposed in \cite{LopezMonteiro.2021} to fairly power a set of EH IoT devices.  
	Note that average CSI is much less prone to estimation errors than instantaneous CSI. But more importantly,
	it varies over a much larger time scale, thus it does not require frequent CSI updates, which allows saving important energy resources. Moreover, an average CSI based design still allows attaining important performance gains since WET channels are typically under strong  line-of-sight (LOS) influence, and consequently have strong deterministic components. 	In fact, it was shown that the EH system performance approaches that under a full CSI precoder when channels are under favorable propagation conditions. 
	However, although a statistical WET precoding may be viable in quasi-static deployments, some energy is still needed to learn the main channel features, which may be 
	still unaffordable for ultra-low power devices. This, combined with the need to power more dynamic deployments, urged the development of CSI-free designs. For instance, authors in \cite{Clerckx.2018} proposed a CSI-free method, herein referred as ``random phase sweeping with energy modulation/waveform'' (RPS-EMW), relying on multiple dumb antennas transmitting phase-shifted signals to induce fast fluctuations on a slow-fading wireless channel. However, transmit diversity is not fully realized, and no average harvesting gains over single antenna WET are reachable with RPS-EMW. Meanwhile, full transmit diversity, but again no average harvesting gains, can be attained  with the  ``switching antennas'' (SA)  and ``all antennas transmitting independent signals'' (AA-IS) CSI-free schemes  proposed in \cite{LopezAlves.2019} and \cite{LopezMontejo.2020}, respectively. By considering the non-linearity of the EH receiver, it is demonstrated in \cite{LopezMontejo.2020} that those devices far from the PB benefit slightly more from SA than from AA-IS, while those closer to the PB benefit slightly more from the latter. On the other hand, the ``all antennas transmitting the same signal'' (AA-SS) CSI-free scheme analyzed in \cite{LopezAlves.2019} does provide average EH gains, and it was further optimized in \cite{LopezMontejo.2020} to provide wider coverage. Thus, we refer to the original \cite{LopezAlves.2019} and optimized \cite{LopezMontejo.2020} proposals as AA-SS-I, and AA-SS-II, respectively. However, these schemes are only capable of serving specific clustered IoT deployments depending on the orientation of the PB's antenna array. To overcome this rigidity and attain higher performance gains, hybrid (between SA, AA-IS, AA-SS-I and AA-SS-II) schemes possibly combined with  deployment-specific antenna orientation were also conceptually put forward in \cite{LopezMontejo.2020}.
	
	Here, we take a very important step forward by proposing a novel CSI-free WET scheme that exploits antenna rotation to outperform its above state-of-the-art counterparts, and even approach quickly (and surpass) the performance of a traditional full-CSI strategy as the number of served devices increases. Our specific contributions are the following: 
	\begin{itemize}
	 \item we propose the novel  ``rotary antenna beamforming'' (RAB) CSI-free scheme, which re-uses the precoder for AA-SS-II in \cite{LopezMontejo.2020}, but herein we fully  realize its potential by including a rotor into the PB's uniform linear array (ULA) 	 
	 while making the array continuously rotate.\footnote{Rotary antenna arrays have been already proposed in the literature for different scenarios/applications. See our discussions on this later in Section~\ref{rotation}.}  
	 By doing this, an average EH gain of approximately $0.85\sqrt{M}$, where $M$ is the number of PB's antenna elements, over a single-antenna transmission (or over the performance of a CSI-free scheme such as SA, AA-IS and RPS-EMW) is attainable. Moreover, this gain is independent of the devices' positions since the resulting effective radiation pattern of RAB is omnidirectional;
	 \item to avoid the use of a costly rotor equipment and ease the implementation,  we propose using a stepper rotor with at least $M$ equally-spaced steps, and show that the resulting effective radiation pattern becomes quasi-omnidirectional;
	 \item we optimize RAB performance by means of a rotation-specific power control mechanism exploiting devices' positioning information. The optimum power control comes from solving a linear program (LP) \cite{Vanderbei.2015}, thus, can be solved efficiently, although not in closed-form. To overcome the latter issue, an alternative simpler power control mechanism that allows reaching near-optimum performance is provided in closed-form;
	 \item we include specific absorption rate (SAR) constraints to make  RAB compliant with electromagnetic field (EMF) regulations, thus, avoiding potential risks to human health. Numerical results illustrate that RAB is less sensitive to such constraints than traditional CSI-based precoders; 
	 \item  we show numerically that RAB outperforms all the state-of-the-art CSI-free WET schemes available in the literature. Even without optimized power control, its performance may be superior to that provided by a full-CSI precoder when the number of devices grows larger than the number of PB's antenna elements. Moreover, we show that by adopting the proposed power control, RAB can easily outperform the full-CSI precoder under full LOS conditions. However, a performance degradation occurs when channels are under the influence of some non-LOS (NLOS) components, for which the full-CSI precoder may slightly outperform RAB;
	 \item we discuss some important practicalities related to RAB. Specifically, we comment on the robustness of RAB against NLOS compared to other CSI-free WET schemes, and delve into the possibility of generalizing RAB implementation to scenarios with PBs using different antenna array topologies.   
	\end{itemize}
    Note that the main analysis and discussions carried out in the paper are for LOS scenarios, for which the performance under the full CSI precoder  matches that of the optimal statistical precoder in \cite{LopezMonteiro.2021}. 
    	\begin{table}[t!]
    	\centering
    	\caption{List of Acronyms}
    	\label{table_0}
    	\begin{tabular}{l  l}
    		\hline
    		\textbf{Acronym} & \textbf{Definition} \\
    		\hline
    		AA-IS & all antennas transmitting independent signals \\
    		AA-SS & all antennas transmitting the same signal \\
    		BS & base station \\
    		CSCG & circularly-symmetric complex Gaussian \\
    		CSI & channel state information \\
    		EH & energy harvesting \\
    		EMF & electromagnetic field \\
    		FCC & Federal Communications Commission \\
    		IoT & Internet of Things \\
    		LOS & line-of-sight \\
    		LP & linear program \\
    		MIMO & multiple-input multiple-output \\
    		MISO & multiple-input single-output \\
    		MPE & maximum permissible exposure\\
    		NLOS & non-LOS \\
    		PB & power beacon \\ 
    		RAB & rotary antenna beamforming \\   	
    		RF & radio frequency \\
    		RPS-EMW\!\! & random phase sweeping with energy modulation/waveform\\
    		SA & switching antennas \\
    		SAR & specific absorption rate \\
    		SDP & semi-definite program \\
    		ULA & uniform linear array \\
    		WET & wireless energy transfer\\		
    		\hline
    	\end{tabular}
    \end{table}

	The remainder of this article is organized as follows. Section~\ref{system} presents the system model and overviews the state-of-the-art CSI-free WET schemes. Section~\ref{bar} introduces and discusses the proposed RAB scheme, which is further optimized via proper power control mechanisms in Section~\ref{optimization}. Section~\ref{practical} reveals important practical considerations, while Section~\ref{results} discusses numerical performance results, and Section~\ref{conclusions} concludes the article. Tables~\ref{table_0} and ~\ref{table_1} list the acronyms and main symbols used throughout the paper.
	\newline\textit{Notation:} 
Boldface lowercase/uppercase letters denote column vectors/matrices. For instance, $\mathbf{x}=\{x_i\}$, where $x_i$ is the $i^\text{th}$ element of vector $\mathbf{x}$, and as a special case $\mathbf{1}$ denotes a vector of ones; while $\mathbf{X}=\{X_{i,j}\}$, where $X_{i,j}$ is the $i^\text{th}$ row $j^\text{th}$ column element of matrix $\mathbf{X}$. Superscripts $(\cdot)^T$ and $(\cdot)^H$ denote the transpose and Hermitian operations, respectively, while $\Tr(\cdot)$ represents the trace of a matrix.
The curled inequality symbol $\succeq$ represents component-wise inequality between vectors.  
Meanwhile, $\mathbb{C}$ is the set of complex numbers, and $\mathbbm{i}=\sqrt{-1}$ is the imaginary unit. Additionally, $|\cdot|$ is the absolute (or cardinality when applied to sets) operation, while  $\mathrm{mod}(a, b)$ is the modulo operation, which provides the remainder of the integer division of $a$ by $b$. $\mathbb{E}[\!\!\ \cdot\ \!\!]$ denotes expectation, $f_Y(y)$ denotes
the probability distribution function of random variable $Y$, and $\mathbf{w}\sim\mathcal{CN}(\mathbf{0},\mathbf{R})$ is a circularly-symmetric complex Gaussian (CSCG) random vector with zero mean and covariance matrix $\mathbf{R}$. Finally, $J_0(\cdot)$ denotes the Bessel function of the first kind and order $0$ \cite[Sec. 10.2]{Thompson.2011}.
	\begin{table}[t!]
	\centering
	\caption{List of Symbols}
	\label{table_1}
	\begin{tabular}{l  l}
		\hline
		\textbf{Symbol} & \textbf{Definition} \\
		\hline
		$\mathcal{S}$, $S_i$  & set of EH devices, and $i^\text{th}$ EH device in the set\\
		$\theta_i$ & $S_i$' azimuth angle relative to PB's ULA boresight\\
		$\mathbf{h}_i$ & normalized LOS channel between PB's ULA and $S_i$\\
		$M$  & number of transmit antenna elements at the PB \\	
		$\phi_{j,i}$ & relative LOS phase shift of the $j^\text{th}$ antenna element of $S_i$\\
		$x_k, y_i$ & $k^\text{th}$ PB's transmit signal, and signal received at $S_i$\\
		$\beta_i$ & path loss between PB's ULA and $S_i$\\
		$K$ & number of energy signals\\
		$\mathbf{v}_k$ & normalized precoding vector associated to $k^\text{th}$ energy signal\\
		$p_{T,k}$ & transmit power of $k^\text{th}$ energy signal\\
		$p_{T}$ & total transmit power of the PB\\
		$P_i$, $\bar{P}_i$  & incident (average) RF power available at $S_i$\\
		$g(\cdot),\eta,E_i$  & EH transfer function, EH efficiency, and harvested energy\\
		$G,G^\text{ric}$ & PB's radiation pattern, power gain (over an isotropic
		radiator)\\
		& in pure LOS and Rician fading channels, respectively\\
		$\tilde{\mathbf{h}}_i$ & Rician NLOS channel component between PB's ULA and $S_i$\\
		$\mathbf{h}^\text{ric}_i$ & Rician fading channel between PB's ULA and $S_i$\\
		$\mathbf{V}$ & transmit covariance matrix\\
		$\mathbf{H}^\text{ric}_i$, $\mathbf{R}_i$ & Rician/NLOS channel covariance matrix\\
		$\bar{G}$, $\bar{G}_i$ & PB's average RF power gain over an isotropic radiator,\\
		 & and that experienced by $S_i$ \\
		 $d_i$ & distance between $S_i$ and the PB\\
		 $\alpha$ & path-loss exponent\\
		 $p_j$ & PB's transmit power at the $j^\text{th}$ angular rotation\\
		 $\xi$ & minimum RF energy available at every EH device\\
		 $I_j$ & set of indexes of all the EH devices served by the main \\
		 & energy beams at the $j^\text{j}$ ULA angular rotation\\
		 $\text{SAR}_{j,z}$ & $z^\text{th}$ SAR measure corresponding to $j^\text{th}$ angular rotation\\
		 $\mathbf{S}^{(z)}$ & $z^\text{th}$ SAR measurement matrix\\
		 $\delta_z$ & $z^\text{th}$ SAR constraint\\
		\hline
	\end{tabular}
\end{table}

	\section{System Model}\label{system}
\begin{figure}[t!]
   	\centering
   	\includegraphics[width=0.42\textwidth]{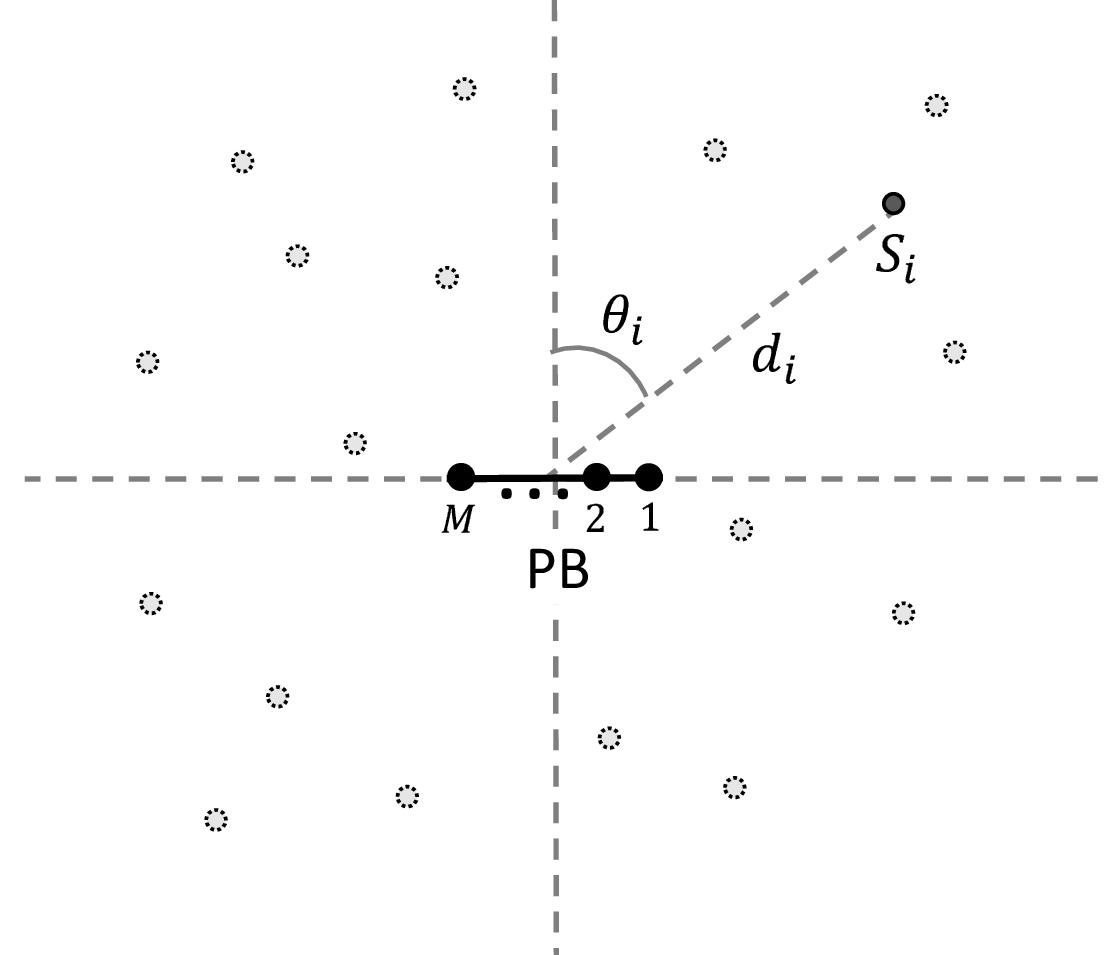}
   	\caption{System model: PB equipped with ULA powering nearby (randomly uniformly distributed) EH devices.}
   	\label{Fig1}
   \end{figure}

	Consider a PB equipped with a ULA of $M$ half-wavelength spaced antenna elements. The PB powers wirelessly a massive set $\mathcal{S}=\{S_i\}$ of single-antenna sensor nodes uniformly randomly deployed around it such that $f_{\theta_i}(\theta_i)=\tfrac{1}{2\pi}$, where $\theta_i\in[0,2\pi]$ is the azimuth angle relative to the boresight of the transmitting ULA as illustrated in Fig.~\ref{Fig1}.\footnote{Note that the assumption of a uniform distribution deployment  refers only to the angular domain, while the distance from the PB to the EH devices may follow an arbitrary distribution.} We consider a LOS channel model, which
	is valid under short distance operation as usual in WET systems \cite{Lopez.2021,Dai.2018,LopezMonteiro.2021,Clerckx.2018,LopezMontejo.2020,Zhang.2020}. However, later on we briefly discuss the impact of small-scale fading and NLOS conditions. 
	
	The normalized LOS (geometric) channel between PB's ULA and $S_i$ is given by \cite[Ch.~5]{Hampton.2014}
	\begin{align}
	\mathbf{h}_i = e^{\mathbbm{i}\varphi_0}[1,e^{\mathbbm{i}\phi_{1,i}},e^{\mathbbm{i}\phi_{2,i}},\cdots,e^{\mathbbm{i}\phi_{M-1,i}}]^T,\label{hi}
	\end{align}
	where $\varphi_0$ accounts for an initial phase shift which depends on the link distance $d_i$ between the PB and $S_i$, but since it impacts all antenna elements equally we can ignore it without loss of generality \cite{Hampton.2014,LopezMontejo.2020}, while
	\begin{align}
	\phi_{j,i} = -j\pi\sin \theta_i,\qquad j=0,1,\cdots,M-1.
	\end{align}
	Additionally, the power path-loss is denoted as $\beta_i$ and we assume that the user-specific downlink WET channels are not known at the PB.		
   \subsection{Signal Model}\label{signal}
   The PB transmits $K$ energy symbols $x_{k}\in\mathbb{C}$ per channel use such that the RF signal received by $S_i$ is given by
   \begin{align}
   y_i = \sum_{k=1}^{K}\sqrt{p_{T,k}\beta_i} \mathbf{v}_{k}^T\mathbf{h}_{i}x_{k},\label{yi}
   \end{align}
   where $\mathbf{v}_{k}\in\mathbb{C}^{M}$ is the normalized precoding vector associated to $x_{k}$, $K\le M$, and $p_{T,k}$ is its corresponding transmit power where $\sum_{k=1}^{K}p_{T,k}=p_T$. Note that noise impact is ignored since it is practically null for EH purposes \cite{Kashyap.2016,Khan.2018,Zeng.2015,Wang.2016,Dai.2018,Liang.2019,Rosabal.2020,Wang.2019,LopezMonteiro.2021,Clerckx.2018,LopezAlves.2019}, and it is assumed that symbols $x_{k}$ are independent and identically distributed (i.i.d.) power-normalized random, i.e., $\mathbb{E}[|x_k|^2]=1,\ \mathbb{E}[x_k^Hx_{k'}]=0,\ \forall k\ne k'$. Then, the incident RF power (averaged over the signal waveform/randomness) available at each $S_i$    
   is given by
   \begin{align}
   P_i&=\mathbb{E}_x[|y_i|^2]\nonumber\\
   &\stackrel{(a)}{=}\!\mathbb{E}_x\Bigg[\bigg(\sum_{k=1}^{K}\!\sqrt{p_{T,k}\beta_i} \mathbf{v}_{k}^T\mathbf{h}_{i}x_{k}\bigg)^{\!H}\!\bigg(\sum_{k=1}^{K}\!\sqrt{p_{T,k}\beta_i} \mathbf{v}_{k}^T\mathbf{h}_{i}x_{k}\bigg)\Bigg] \nonumber\\
   &\stackrel{(b)}{=}\beta_i\sum_{k'=1}^K\sum_{k''=1}^K\sqrt{p_{T,k'}p_{T,k''}} (\mathbf{v}_{k'}^T\mathbf{h}_i)^H\mathbf{v}_{k''}^T\mathbf{h}_i\mathbb{E}[x_{k'}^Hx_{k''}]\nonumber\\
   &\stackrel{(c)}{=}\beta_i\sum_{k=1}^{K}p_{T,k}|\mathbf{v}_k^T\mathbf{h}_i|^2, \label{Pk} 
   \end{align}
  where  (a) comes from using \eqref{yi}, (b) follows after re-arranging terms, and (c) is immediately attained from leveraging the assumption of independent and power-normalized signals.
   Meanwhile, the energy harvested by $S_i$ converges to $E_i =\mathbb{E}_x[g(|y_i|^2)]$ for sufficiently-long coherence channel intervals,  as typical of quasi-static WET setups.
   Note that $g$ is a real function modeling the relation between the incident RF power and harvested power. Herein, we consider a simple linear EH model, i.e., $g(|y_i|^2)=\eta |y_i|^2$, where $\eta\in[0,1)$,  for analytical tractability, thus facilitating the discussions and deriving important performance insights as in \cite{Liang.2019,Kashyap.2016,Wang.2019,Zeng.2015,LopezAlves.2019}. Note that using the linear model allows us to state that $\mathbb{E}_x[g(|y_i|^2)]=g(\mathbb{E}_x[|y_i|^2])=g(P_i)$, thus the average harvested power increases linearly with the average incident RF power, and we can refer to power and energy indistinctly by normalizing the EH time. In general, maximizing the average harvested energy translates to maximizing the average RF power, and we can focus on the latter.  Although the scaling is not linear in practice as EH non-linearity impacts the performance  \cite{Alevizos.2018,Clerckx.2018}, we let the analysis and related discussions considering an EH non-linear model for future work.
   \subsection{State-of-the-Art CSI-free WET Schemes}\label{csiArt}
   Next, we briefly discuss the main state-of-the-art CSI-free strategies available in the literature, which will be used later in Section~\ref{results} as benchmarks for our proposed scheme.
   \paragraph{AA–IS \cite[Sec. III-B]{LopezMontejo.2020}} 
   The PB transmits signals  independently generated across the antenna elements and with equal power, thus $K=M$, and, e.g., $v_{k,j}=c_{k-j}/\sqrt{M}$, where $c_{0}=1$ and $c_{m}=0$, $\forall m\ne 0$. The resulting radiation pattern is omnidirectional with no average harvesting gains but $M-$fold diversity gain with respect to single antenna transmissions in independent Rician fading channels.
   \paragraph{SA \cite[Sec. III-A]{LopezAlves.2019}} 
   The PB transmits a signal with full power by one antenna at a time such that all antenna elements are used during a coherence time block. The radiation pattern is omnidirectional since only one antenna is active at a time, and the performance provided by SA is shown to be equal/similar to that of AA-IS under a linear/non-linear EH model.
   \paragraph{AA–SS \cite{LopezAlves.2019,LopezMontejo.2020}}
   The PB transmits the same signal simultaneously with all antenna elements, i.e., $K=1$, and with equal power at each. There are two basic AA-SS configurations: 
   \begin{itemize}[leftmargin=*]
 \item AA-SS-I \cite[Sec. III-A]{LopezAlves.2019}, where the PB applies the precoder $\mathbf{v}=M^{-\tfrac{1}{2}}\mathbf{1}$ to attain an energy beam towards the ULA boresight directions;
 \item AA-SS-II \cite[Sec. IV-B]{LopezMontejo.2020}, where the PB applies the precoder
 \begin{align}
 v_j = \frac{1}{\sqrt{M}}e^{\mathrm{mod}(j-1,2)\pi\mathbbm{i}}=\sqrt{\frac{1}{M}}\big[1,-1,1,\cdots\big]^T\label{vj}
 \end{align}
 to attain the energy beams that maximize the RF energy expected to be available at a random device in the area with channels subject to Rician fading. The resulting main beams are wider than in AA-SS-I, and oriented toward $90^\circ$ offset from ULA's boresight directions.
   \end{itemize}
Under both AA-SS schemes, some devices will harvest significantly more energy than others depending on their respective locations with respect to the PB's antenna array.
\paragraph{RPS-EMW \cite[Secs. III \& IV]{Clerckx.2018}}
The PB transmits the same  signal simultaneously through all the antenna elements but with some distinctive features: i) the beamforming phase is taken uniformly and independently distributed over $[0, 2\pi]$; and ii) energy modulation/waveform  is introduced to provide further enhancements by taking advantage of the non-linearity of the EH circuitry. For benchmarking results, we adopt a CSCG waveform in Section~\ref{results}.
   \section{CSI-free Rotary Antenna Beamforming}\label{bar}
   In this section, we present and discuss the performance of our proposed  RAB scheme, which consists of two main design components: i) the CSI-free beamformer (Subsection~\ref{beam}) and ii) ULA rotation and related optimization (Subsection~\ref{rotation}).
   \subsection{Beamformer Design}\label{beam}
\begin{figure}[t!]
   		\centering
   		\includegraphics[width=0.48\textwidth]{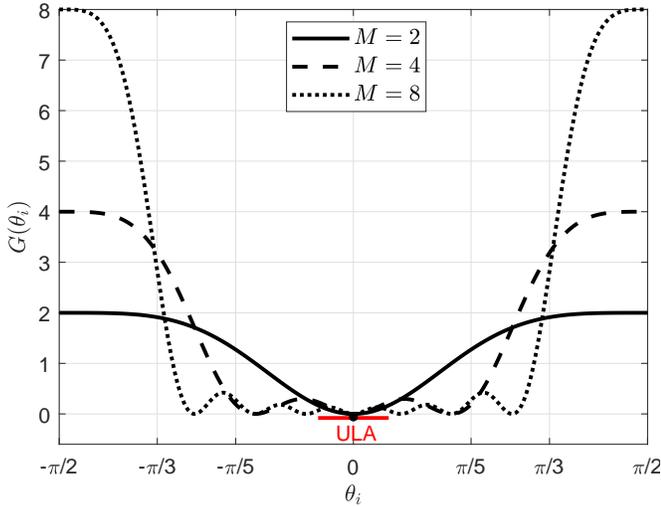}
   		\caption{Radiation pattern, $G(\theta_i)$ vs $\theta_i$, when using \eqref{vj}.}
   		\label{Fig2}
   	\end{figure}

   First, note that the PB's radiation pattern over a certain angular domain can be derived from \eqref{Pk} by removing the large scale channel effect and PB's transmit power contribution as 
   \begin{align}
   G(\theta_i,\mathbf{v}_k) = \frac{P_i(\theta_i,\mathbf{v}_k)}{p_T\beta_i}.\label{Gi}
   \end{align}  
In fact, $G(\theta_i,\mathbf{v}_k)$ represents the power gain (over an isotropic radiator) in the direction $\theta_i$ when using the normalized precoder $\mathbf{v}_k$. 
   Now, we propose using the single-signal beamformer leading to the best radiation pattern in terms of average RF energy available along the entire angular domain, i.e.,
   \begin{align}
   \mathbf{v}_k^\text{opt}=\arg\max_{\mathbf{v}_k}\int_{0}^{2\pi}G(\theta_i,\mathbf{v}_k)\mathrm{d}\theta_i
   \end{align} 
   In \cite{LopezMontejo.2020}, it is shown that beamformer \eqref{vj} fulfills this criterion in case of Rician fading channels (after averaging over the channel realizations), for which the normalized fading channel vector can be written as
   \begin{align}
   \mathbf{h}_i^\text{ric} = \mathbf{h}_i + \tilde{\mathbf{h}}_i, \label{hric}
   \end{align}
   where $\mathbf{h}_i$ is the LOS channel given in \eqref{hi}, and $\tilde{\mathbf{h}}_i\sim \mathcal{CN}(\mathbf{0},\mathbf{R}_i)$ is the NLOS channel component. Plugging \eqref{hric} into $\mathbf{h}_i$ in \eqref{Pk} with $K=1$ (and avoiding sub-index $k$), while then using \eqref{Gi} and taking the expectation over the channel realizations, yields
   \begin{align}
   \mathbb{E}\big[G^\text{ric}(\theta_i,\mathbf{v}_k)\big]&= \mathbb{E}\big[|\mathbf{v}^T\mathbf{h}_i^\text{ric}|^2\big]=\mathbb{E}\big[(\mathbf{v}^T\mathbf{h}_i^\text{ric})^H(\mathbf{v}^T\mathbf{h}_i^\text{ric})\big]\nonumber\\
   &=\mathbb{E}\big[\Tr(\mathbf{V}^T\mathbf{H}_i^\text{ric})\big], \label{eqV}
   \end{align}
   where $\mathbf{V}=\mathbf{v}\mathbf{v}^H$ and $\mathbf{H}_i^\text{ric}=\mathbf{h}_i^\text{ric}\mathbf{h}_i^{\text{ric} H}$. Authors in \cite{LopezMontejo.2020} demonstrated that $\mathbf{v}$ constructed as in \eqref{vj} maximizes $\int_{0}^{2\pi}\mathbb{E}[G^\text{ric}(\theta_i,\mathbf{v}_k)]\mathrm{d}\theta_i$.    
   Since $\mathbf{v}$ in \eqref{vj} does not depend on the instantaneous channel realizations,  we can further reduce \eqref{eqV} by using \eqref{hric} as follows
   \begin{align}
   \mathbb{E}[G^\text{ric}(\theta_i,\mathbf{v}_k)]&=\Tr\big(\mathbf{V}^T\mathbb{E}[\mathbf{H}_i^\text{ric}]\big)\nonumber\\
   &=\Tr\big(\mathbf{V}^T\mathbf{h}_i\mathbf{h}_i^H\big) + \Tr\big(\mathbf{V}^T\mathbf{R}_i\big),\label{pir}
   \end{align}
   since $\mathbb{E}[\mathbf{H}_i^\text{ric}]=\mathbb{E}[\mathbf{h}_i\mathbf{h}_i^H+\mathbf{h}_i\tilde{\mathbf{h}}_i^H+\tilde{\mathbf{h}}_i\mathbf{h}_i^H+\tilde{\mathbf{h}}_i\tilde{\mathbf{h}}_i^H]=\mathbf{h}_i\mathbf{h}_i^H+\mathbf{R}_i$.
   Now, observe that since $\mathbf{v}$ in \eqref{vj} maximizes \eqref{pir}, and does not depend on the entries of $\mathbf{R}_i$, it follows that it also maximizes $\int_{0}^{2\pi}\Tr\big(\mathbf{V}^T\mathbf{h}_i\mathbf{h}_i^H\big)\mathrm{d}\theta_i$, which is the component related to the LOS.   
    As such, we propose here using \eqref{vj} as well. Thus, by substituting \eqref{hi} and first formulation of $\mathbf{v}$ in \eqref{vj} into \eqref{Pk} and \eqref{Gi}, we attain
   \begin{align}
   G(\theta_i)&=G(\theta_i,\mathbf{v}_k)\big|_{\mathbf{v}_k\text{ as in \eqref{vj}}}\nonumber\\
   &=\frac{1}{M}\bigg|\sum_{j=1}^{M}e^{\mathbbm{i}(\mathrm{mod}(j-1,2)\pi-(j-1)\pi\sin\theta_i)}\bigg|^2\nonumber\\
   &\stackrel{(a)}{=}\frac{1}{ M}\bigg[\bigg(\sum_{j=1}^{M}\!\cos\!\big(\mathrm{mod}(j\!-\!1,2)\pi\!-\!(j\!-\!1)\pi\sin\theta_i\big)\!\bigg)^2\nonumber\\
   &\qquad +\bigg(\sum_{j=1}^{M}\sin\big(\mathrm{mod}(j\!-\!1,2)\pi\!-\!(j\!-\!1)\pi\sin\theta_i\big)\bigg)^2\bigg]\nonumber\\
   &\stackrel{(b)}{=}1\!+\!\frac{2}{M}\!\sum_{j=1}^{M-1}\sum_{l=j+1}^M\!\cos\!\big(\mathrm{mod}(j\!-\!1,2)\pi\!-\!(j\!-\!1)\pi\sin\theta_i\nonumber\\
   &\qquad\qquad\qquad - \!   \mathrm{mod}(l\!-\!1,2)\pi\!+\!(l\!-\!1)\pi\sin\theta_i\big)\nonumber\\
   &\stackrel{(c)}{=}1+\frac{2}{M}\sum_{j=1}^{M-1}\sum_{l=j+1}^M\cos\big(\mathrm{mod}(j-l,2)\pi\nonumber\\
   &\qquad\qquad\qquad+(l-j)\pi\sin\theta_i\big),  \label{Pi}
   \end{align}
   where $(a)$ comes from using $e^{\mathbbm{i}x}=\cos(x)+\mathbbm{i}\sin(x)$ and $|x+\mathbbm{i}y|^2=x^2+y^2$, $(b)$ follows from expanding the power and using $\cos(x)^2+\sin(x)^2=1$ and $\cos(x)\cos(y)+\sin(x)\sin(y)=\cos(x-y)$, $(c)$ follows from carrying out simple algebraic operations including modular arithmetic.       
   The resulting radiation pattern is illustrated in Fig.~\ref{Fig2}. Observe that the maxima occur at $90^\circ$ offset from ULA's boresight directions as claimed in \cite{LopezMontejo.2020}, and that the number of minima matches $M-1$ in $\theta_i\in[-\pi/2,\pi/2]$.
   
   Now, the average RF power gain, is given by $\tfrac{1}{2\pi}\int_{0}^{2\pi}G(\theta_i)\mathrm{d}\theta_i$, which using \eqref{Pi} transforms to
   \begin{align}
   \bar{G}  &=1\!+\!\frac{1}{\pi M}\!\sum_{j=1}^{M-1}\sum_{l=j+1}^M\!\int\limits_{0}^{2\pi}\!\!\cos\!\big(\mathrm{mod}(j\!-\!l,2)\pi\!+\!(l\!-\!j)\pi\sin\theta\big)\mathrm{d}\theta\nonumber\\
&   \stackrel{(a)}{=}1+\frac{2}{ M}\sum_{j=1}^{M-1}\sum_{l=j+1}^M(-1)^{l-j}J_0((l-j)\pi)\nonumber\\
&   \stackrel{(b)}{=}1+\frac{2}{ M}\sum_{j=1}^{M-1}(-1)^{j}(M-j)J_0(j\pi), \label{J0}
   \end{align}
   where $(a)$ follows from using the integral representation of $J_0(\cdot)$ \cite[eq. (10.9.1)]{Thompson.2011}, and $(b)$ from a simple structure analysis and simplification of the double summation in $(a)$.      
   
\begin{figure}[t!]
	\centering
	\includegraphics[width=0.48\textwidth]{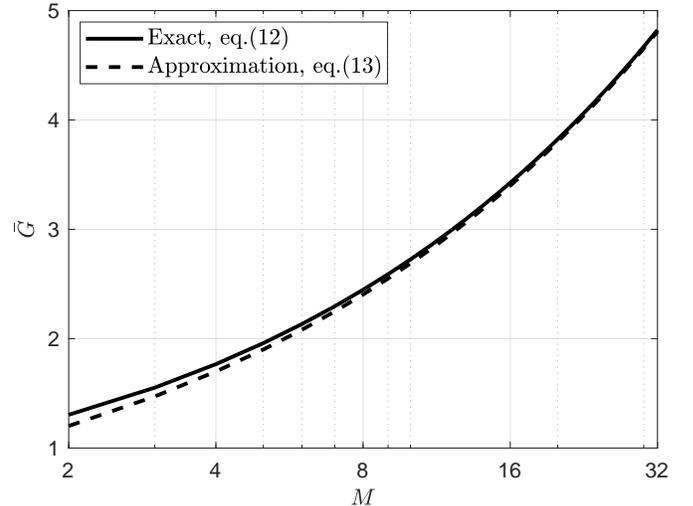}
	\caption{$\bar{G}$ as a function of $M$.}
	\label{Fig3}
\end{figure}

   Note that $\bar{G}$ represents the average EH gain over single antenna WET, and also over state-of-the-art CSI-free omnidirectional WET schemes such as AA-IS, SA and RPS-EMW, for which $\bar{G}=1$. Since $J_0(j\pi)$ is positive/negative for $j$ being even/odd, we have that $\bar{G}>1, \forall M$, thus, one can conclude that using $\mathbf{v}$ in \eqref{vj} leads to a performance superior than that of single antenna, AA-IS, SA and RPS-EMW WET schemes. 
   Besides that, and although \eqref{J0} is in closed form, one cannot get much more insights from it. However, it can be shown that 
   \begin{align}
   \bar{G} \approx 0.85\sqrt{M},\label{avP}
   \end{align}
   as illustrated in Fig.~\ref{Fig3}, which evidences a more explicit relation between $\bar{G}$ and $M$, i.e., a square-root law dependence. This result was obtained by applying standard curve-fitting techniques, and although it is shown only for $M\le 32$, it also holds  tight for $M>32$. Finally, the average RF power gain when using AA-SS-I is given by $\bar{G}\approx 0.64$, which can be derived by following the same previous procedure but using $\mathbf{v}=M^{-\tfrac{1}{2}}\mathbf{1}$ when evaluating \eqref{Pi}. These results are summarized in Table~\ref{table} and evidence the relevance of the CSI-free precoder given in \eqref{vj} since it is the only one that provides average RF power gains that even increase with the number of transmit antenna elements.
  \subsection{PB's ULA Rotation}\label{rotation}
   The average RF power availability in \eqref{avP} can only be attained when averaging over all devices' positions. In fact, devices positioned in the directions of the minima of the radiation pattern in Fig.~\ref{Fig2} will collect little, if any, energy when using the beamformer \eqref{vj} independently of the number of antenna elements. Thus, the simple adoption of the beamformer design previously discussed leads to a rather unfair WET.    
   To overcome this, herein we propose equipping the PB with a rotary-motor and letting it continuously rotate. By doing this, each EH device’s angular position will constantly change with respect to the ULA boresight direction, thus, allowing it to experience the average EH gains claimed in \eqref{J0} and \eqref{avP}. 
   
 The idea of rotary antenna arrays is not new. In \cite{Milligan.1999}, Milligan expanded the idea of using Euler rotation angles for calculating antenna patterns, previously  discussed by Burger in \cite{Burger.1995}, to include physical antenna rotations. Meanwhile, more recently, some potential gains from exploiting adjustable antenna orientation as an extra degree of freedom in scenarios under the influence of some LOS have been explored in \cite{Lysejko.2018,Do.2020,LopezMontejo.2020,LopezMonteiro.2021}. Specifically, the idea of a link-quality specific rotation configuration, where the transmit array could also tune its transmission beam pattern, was patented in  \cite{Lysejko.2018}; while authors in \cite{Do.2020} explored the signal-to-noise gains coming from  configurable transmit/receive ULA. While in the context of WET, authors in \cite{LopezMontejo.2020,LopezMonteiro.2021} have given some initial clues about possible gains from proper PB's ULA orientation when using CSI-limited/free strategies, but a specific rotatory strategy and associated performance gains have not been defined and quantified.
 \begin{table}[!t]
	\centering
	\caption{Average RF Power Gains of CSI-free WET Schemes \cite{LopezAlves.2019,LopezMontejo.2020,Clerckx.2018}}
	\begin{tabular}{lcccccc}
		\toprule
		& AA-IS & SA & RPS-EMW & AA-SS-I & AA-SS-II  \\
		\textbf{$\bar{G}$} & 1 & 1 & 1 & 0.64 & $0.85\sqrt{M}$  \\
		\bottomrule
	\end{tabular}\label{table}
\end{table}

Back to the scenario under discussion in this paper, note that no fully-digital (without mechanical rotation) CSI-free scheme consisting of a combination of several digital precoders (e.g., using spatial multiplexing, or time-specific precoding mimicking a kind of ULA `virtual/digital' rotation) can provide the omnidirectional gains claimed in \eqref{J0} and \eqref{avP}. The reason is that such a fully-digital design would require using precoders that individually perform worse than \eqref{vj}. Thus, the average RF energy gain will be unavoidably smaller than that in \eqref{J0} and \eqref{avP}. Meanwhile, observe  that the proposed scheme may lead to a considerably increase in the PB power consumption since the energy consumed in the mechanical rotations may easily exceed the transmit energy. However, it is more EMF-friendly than its  state-of-the-art CSI-free counterparts, which require increasing the transmit power, thus the electromagnetic pollution, to attain similar performance gains. 
    
\subsubsection{Low-Complexity Rotation}    
\begin{figure}[t!]
	\centering
	\includegraphics[width=0.48\textwidth]{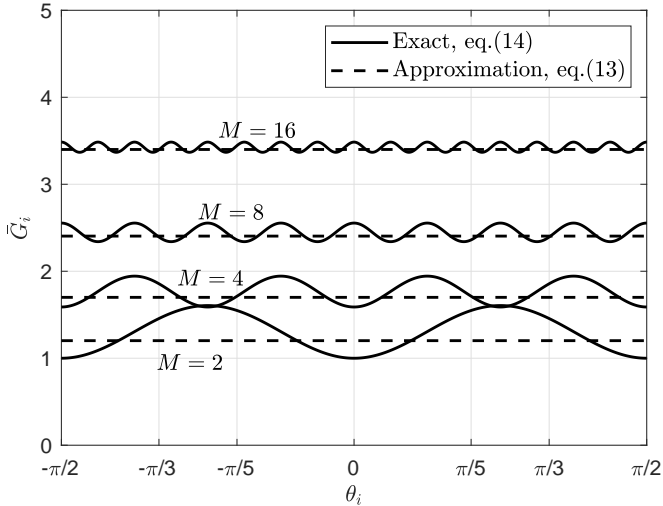}
	\caption{$\bar{G}_i$ as a function of $\theta_i$ for $M\in\{2,4,8,16\}$.}
	\label{Fig4}
\end{figure}

   In practice, a smooth PB's antenna array angular rotation may be difficult to achieve, but it may not even be necessary. 
   Since the number of minima of the radiation pattern was shown to be $M-1$ in the angular domain $[-\pi/2,\ \pi/2]$, we may conclude that at least $M$ angular rotations that cover such angular domain are needed.\footnote{Even just half of such steps are enough by taking advantage of ULA's radiation patterns symmetry and restricting the angular rotation to the domain $[0,\ \pi/2]$. Alternatively, a maximum of $2M$ rotation steps may be needed if completely circular rotations are implemented.}
  Indeed, equipping the PB with a stepper motor with $M$ equally-spaced steps provides already a sufficiently smooth performance, thus, we adopt such spacing here. As a consequence, a certain device $S_i$ will experience an incident average RF power gain (compared to the scenario with an isotropic PB radiator)  given by
   \begin{align}
   \bar{G}_i = \frac{1}{M}\sum_{j=1}^{M}G\Big(\theta_i+\frac{j\pi}{M}\Big),\label{barPi}
   \end{align}
and the resulting performance is illustrated in Fig.~\ref{Fig4}. Note that $\theta_i$ is just set at the beginning prior to any rotation and remains unchanged.
Therefore, $\bar{G}_i$ as a function of $\theta_i$ can be interpreted as the average radiation pattern of the PB's ULA when using RAB with $M$ equally-spaced steps in the angular domain $[-\pi/2,\ \pi/2]$. As evidenced, the radiation pattern is quasi-omnidirectional. 

At this point we would like to note that with the continuous and rapid advances on motor/rotor miniaturization \cite{Beretta.2020}, and the foreseen performance gains \cite{Lysejko.2018,Do.2020,LopezMontejo.2020,LopezMonteiro.2021}, rotary antenna mechanisms may become attractive for several applications in a near future, out of which CSI-free massive WET, as discussed here, is definitely one.
\section{RAB Optimization}\label{optimization}
In this section, we discuss some techniques for optimizing further the performance of our proposed RAB scheme. Specifically, in Subsection~\ref{position}, we design a power control mechanism that exploits devices' positioning information, while in Subsection~\ref{EMF}, we introduce further optimization constraints related to EMF exposure  and discuss how to address them.
\subsection{Exploiting Devices' Positioning Information}\label{position}
In general, positioning information, which may be available in case of static, quasi-static, or even more dynamic IoT deployments, e.g., using energy-efficient localization techniques \cite{Shit.2018,Want.2018,Li.2021}, could be used to improve the WET efficiency \cite{Lopez.2021}. In the scenario under discussion, the devices' positions determine the LOS channel \eqref{hi}. However, one may not directly use the location information to make strong claims regarding the true channel realizations. Thus, one may use \eqref{hi} for performance analysis, or to influence a mechanism design in a statistic (average) sense as done previously in Section~\ref{bar}, but one should not use  \eqref{hi}, for instance, to design an instantaneous CSI-based precoder. The reasons are that i) positioning information is subject to inherent estimation errors, ii) there are always NLOS and channel fluctuation components, and iii) model in \eqref{hi} is also subject to errors coming from hardware/antenna inhomogeneity.

With available positioning information about devices deployment, RAB performance can be improved further. For instance, ULA angular positions pointing to obstructed directions can be avoided, thus, saving energy to improve WET towards the remaining directions. Moreover, it might happen that certain angular directions cover devices much farther than others. In such a case, the time/power allocation to each angular direction should be adjusted accordingly. Let's explore this in more detail next, but  notice first that from the perspective of a linear EH model, optimizing over time or power is indistinguishable, thus we can focus on the latter without loss of generality.\footnote{The adoption of non-linear EH models, where an optimization over time or power substantially differ, is left for future work.}

Let $d_i$ denote the distance between $S_i$ and the PB. Thus, we have that $\beta_i\varpropto d_i^{-\alpha}$, where $\alpha$ is the path-loss exponent, which we assume known beforehand, e.g., as a result of prior measurement studies. Let $p_j$ denote the transmit power of the PB at the $j^\text{th}$ angular rotation, and $\mathbf{p}=[p_1,p_2,\cdots,p_M]^T$, thus, $\mathbf{1}^T\mathbf{p}=p_T$. Then, the average (over all antenna array rotations) RF power available at $S_i$ can be obtained using \eqref{barPi} as 
\begin{align}
\bar{P}_i(\mathbf{p}) \varpropto \frac{d_i^{-\alpha}}{M}\sum_{j=1}^{M}p_jG\Big(\theta_i+\frac{j\pi}{M}\Big)=\mathbf{a}_i^T\mathbf{p},\label{barPij}
\end{align}
where 
\begin{align}
a_{i,j}=\frac{d_i^{-\alpha}}{M}G\Big(\theta_i+\frac{j\pi}{M}\Big).
\end{align}
Then, the optimum power allocation per ULA angular position in terms of WET fairness, $\mathbf{p}^\text{opt}$, comes from solving $\mathrm{maximize}_{\mathbf{p}} \min_i\{\mathbf{a}_i^T\mathbf{p}\}$ subject to $\mathbf{1}^T\mathbf{p}\le p_T$. This problem can be equivalently and straightforwardly transformed to 
\begin{subequations}\label{P}
	\begin{alignat}{2}
	\mathbf{P:}\ \ &\underset{\mathbf{p},\ \xi}{\mathrm{maximize}}       &\ \ \ & 
	\xi \label{P:a}\\ \vspace{-1mm}
	&\text{subject to} & &\   \mathbf{a}_i^T\mathbf{p}\ge \xi,\qquad \forall i,  \label{P:b}\\
	&  & &\ \mathbf{1}^T\mathbf{p}\le p_T,  \label{P:c}
	\end{alignat}	
\end{subequations}
where $\xi$ is an auxiliary optimization variable, which also denotes the minimum RF energy that is available for every device $S_i$ in the network. Then, since \eqref{P} is an LP, it can be optimally solved using standard optimization techniques and tools, e.g., CVX \cite{CVX.2014}, and \texttt{linprog} from MatLab \cite{MATLAB.2018}, with worst-case polynomial complexity \cite{Vanderbei.2015}. 

Although a closed-form solution is difficult to obtain in general, we can design a sub-optimal power allocation as follows. The idea is to exploit the fact that the PB transmit power at each angular rotation must be approximately inversely proportional to the path-loss of the worst-performing device served by the main (strongest) energy beams. The latter are defined to cover the angular domain $\tfrac{j\pi}{M}\pm \tfrac{\pi}{2M}$ ($+\pi$ to account for the ULA symmetry). Let $I_j$ denote the set of indexes of all the devices served by the main energy beams at the $j^\text{th}$ ULA angular rotation, i.e.,
\begin{align}
I_j = \Big\{i\ \Big|\frac{\pi}{2}-\frac{j\pi}{M}-\frac{\pi}{2M} \le \theta_i^*\le \frac{\pi}{2}-\frac{j\pi}{M}+\frac{\pi}{2M}\Big\},
\end{align}
where $\theta_i^* = \theta_i-\pi$ if $\pi/2\le \theta_i\le 3\pi/2$, $\theta_i^*=\theta_i-2\pi$ if $\theta_i\ge 3\pi/2$, otherwise $\theta_i^*=\theta_i$.
Then, $p_j^\text{opt} \appropto \max_{i\in I_j} d_i^{\alpha}$, and using the total power constraint yields
\begin{align}
p_j^\text{opt} \approx \frac{\max_{i\in I_j} d_i^{\alpha}}{\sum_{j=1}^{M}\max_{i\in I_j} d_i^{\alpha}}p_T.\label{popt}
\end{align}
\subsection{Containing the EMF Exposure}\label{EMF}
WET systems must be subject to strict regulations on the EMF radiation levels that users may be exposed to, e.g., in terms of SAR and/or maximum permissible exposure (MPE). These regulations minimize the potential biological effects, e.g., tissue heating, caused by RF radiation. 
Specifically, SAR measures the absorbed power in a unit mass of human tissue by using units of Watt per kilogram [W/kg], while MPE  constraints the level of EMF radiation specified [W/$\mathrm{m}^2$] units. Herein we merely focus on the former, which, for short distances, e.g., centimeters or few meters, becomes more relevant than MPE \cite{Zhang.2020}. 

In case of multi-antenna setups, preliminary results in \cite{Hochwald.2014} have shown that the pointwise SAR can be modeled as a quadratic form of the transmitted signal. Since SAR is a quantity averaged over the transmit signals, it can be expressed as \cite{Zhang.2020}
\begin{align}
\text{SAR}_{j,z}&= p_j\mathbf{v}^H\mathbf{S}^{(z)}\mathbf{v}=\Tr\big(\mathbf{S}^{(z)}\mathbf{V}\big) =p_j s_z \label{SAR1}
\end{align}
for the $j^\text{th}$ ULA angular rotation, where $\mathbf{S}^{(z)}$ is an $M\times M$ positive-definite conjugate-symmetric SAR matrix and its entries have units of W/kg, while $\mathbf{V}=\mathbf{v}\mathbf{v}^H$. 
Moreover, notice that a SAR-compliant system needs to account for various human exposure constraints, e.g., on the whole body, partial body, feet, hands, with various measurement limitations according to  the Federal Communications Commission (FCC) regulations \cite{Chan.2001}. Therefore, multiple measurement matrices $\mathbf{S}^{(z)},\ z=1,2,\cdots$, are generally defined, each with its corresponding SAR constraint $\delta_z$.
Back to \eqref{SAR1}, $s_z$ is defined as
\begin{align}
s_z=\frac{1}{M}\sum_{\forall l,m}(-1)^{\mathrm{mod}(l+m,2)}S_{l,m}^{(z)}.\label{sz}
\end{align}
Observe that the last line of \eqref{SAR1}, together with \eqref{sz}, comes from using the last formulation of $\mathbf{v}$ in \eqref{vj}, which leads to the conclusion that $\Tr(\mathbf{S}^{(z)}\mathbf{V})$ matches the sum of all the elements of matrix $\mathbf{S}^{(z)}$ with indexes $l,m$ such that $l+m$ is even, minus the sum of all the elements of matrix $\mathbf{S}^{(z)}$ with indexes $l,m$ such that $l+m$ is odd. 

Finally, since \eqref{SAR1} depends linearly on $p_j$, we can easily impose SAR constraints of the form
\begin{align}
\text{SAR}_{j,z}&\le \delta_z, \qquad\ \ \! \forall z, \forall j\in\mathcal{J}_\text{sar}\nonumber\\
p_j&\le \min_z \frac{\delta_z}{s_z}, \qquad \forall j\in\mathcal{J}_\text{sar} \label{sar}
\end{align}
to the set $\mathcal{J}_\text{sar}$ of angular rotations with main associated beams pointing to detected humans in the proximity. Constraints \eqref{sar} are linear on $p_j$, thus,
we can easily incorporate them to $\mathbf{P}$ in \eqref{P} since the resulting problem is again an LP and can be solved globally. Some results on SAR-constrained power-control  optimization are drawn in Section~\ref{results} for illustration.
	\section{Practical Considerations}\label{practical}
	Herein, we discuss some practicalities related to the implementation of the proposed RAB scheme.
	\subsection{Robustness Against NLOS}\label{nlos}
\begin{figure}[t!]
		\centering
		\includegraphics[width=0.4\textwidth]{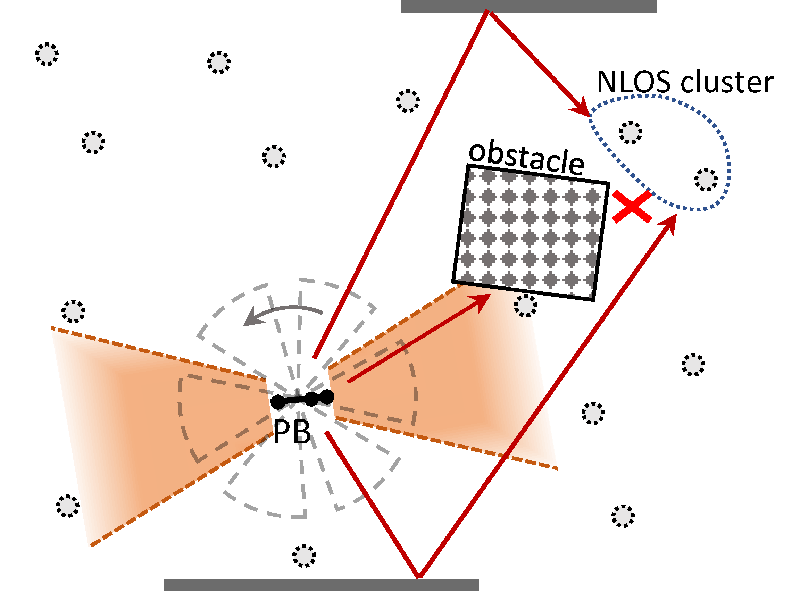}
		\caption{Example of RAB-enabled WET serving clusters of NLOS EH devices.}
		\label{Fig5}
	\end{figure}

	For each angular rotation, RAB creates strong beams toward $90^\circ$ offset from the boresight directions of the PB's ULA. However, as illustrated in Fig.~\ref{Fig5}, it might happen that the LOS towards some devices positioned in such directions is obstructed. 
	Fortunately, the effective RAB's radiation pattern, which is the radiation pattern averaged over all the rotations, is quasi-omnidirectional\footnote{Omnidirectional for a smoothly continuous rotation. In case of using stepper motors, the radiation pattern becomes completely omnidirectional as the number of steps approaches infinity.} as illustrated in Fig.~\ref{Fig4}, with time-specific strong beams pointing to specific spatial directions. Thus, even when the LOS channels towards certain devices may be obstructed, there are strong chances that they can still be served by exploiting strong beam reflections, specially for large $M$. This is because the power of the energy beams, as well as their resolution and the number of rotations, increase with $M$. The latter favors NLOS reachability, while the former makes the energy delivered through the NLOS paths greater.
	\subsection{Generalizing RAB}\label{general}
	Our proposed RAB mechanism can be generalized to fit other antenna array topologies, e.g., planar, circular, cylindrical. The design steps carried out for the ULA case in Section~\ref{bar} can be generalized as follows
	\paragraph{Beamformer Design}
	This phase consists of finding the phase shifts between the transmit antenna elements that provide the best radiation pattern when transmitting the same signal (either multi-sine or modulated waveform) over all the antenna elements. This is done under the assumption of LOS between the PB and the EH devices, which is expected to hold (at least approximately) since WET channels are typically LOS-dominant due to the short distances. To carry this out, one needs to 
	\begin{enumerate}
		\item use the LOS multiple-input single-output (MISO) geometry model, which depends on the PB's antenna array topology, to write the expected available RF energy at a certain spatial direction (specified by zenith and azimuth angles) and as a function of an introduced phase shift (beamforming) vector. Path-loss needs to be ignored/normalized.
		\item	find the phase shift (beamforming) vector that maximizes the average available RF energy along all the spatial directions. To do this, one assumes the possible spatial directions are uniformly random. This uniformly random assumption for the spatial directions needs to be done even in the case that devices' positioning information is known in advance.  
		An optimization using positioning information may be carried out as part of the next step.
	\end{enumerate}
	\paragraph{Array Rotation and Optimization}
	The PB is equipped with a rotary motor that allows to rotate the antenna array to the desired orientation. Based on the radiation pattern that can be generated using the phase shifts designed in the previous step, the PB chooses the rotation configuration and/or power/time allocation that optimizes the RF energy transfer into the desired spatial directions.\footnote{Notice that the performance optimization may be subject to EMF health constraints, e.g., based on SAR as discussed in Section~\ref{EMF}.} Positioning information is useful for this, if available. 
	If devices' positioning information is not available, the rotation is over all the available spatial directions and power/time is equally partitioned. 
	
	In a practical implementation, the rotations may not be smooth since stepper motors may need to be used. In such a case, the number of required motor rotation steps is proportional to the number of PB's antenna elements, and may depend on the array topology. The rotations are performed continuously, or periodically (with a time stay at each step), and the corresponding configuration may be updated in case topological information changes become available. 	
	\section{Numerical Results}\label{results}
	This section presents numerical results on the performance of the proposed CSI-free RAB scheme. We assume a path-loss model with exponent $2$ and a non-distance dependent loss of 40 dB (@2.4 GHz), i.e., $\beta_i = 10^{-4}d_i^{-2}$, while the PB's total transmit power is set to $3$ W, which matches practical PB hardware figures.\footnote{See in \url{https://www.powercastco.com/products/powercaster-transmitter/} for an example of a PB with 1W or 3W effective isotropic radiated power.} The performance of RAB is evaluated assuming the PB is equipped with a stepper rotary motor with $M$ equally-spaced steps as recommended in Section~\ref{rotation}. We resort to CVX \cite{CVX.2014} for optimally solving $\mathbf{P}$ in \eqref{P}, with or without SAR constraints, whenever required, and set $M=4$ unless stated otherwise.
	
	\begin{figure}[t!]
		\centering	
		\includegraphics[width=0.085\textwidth]{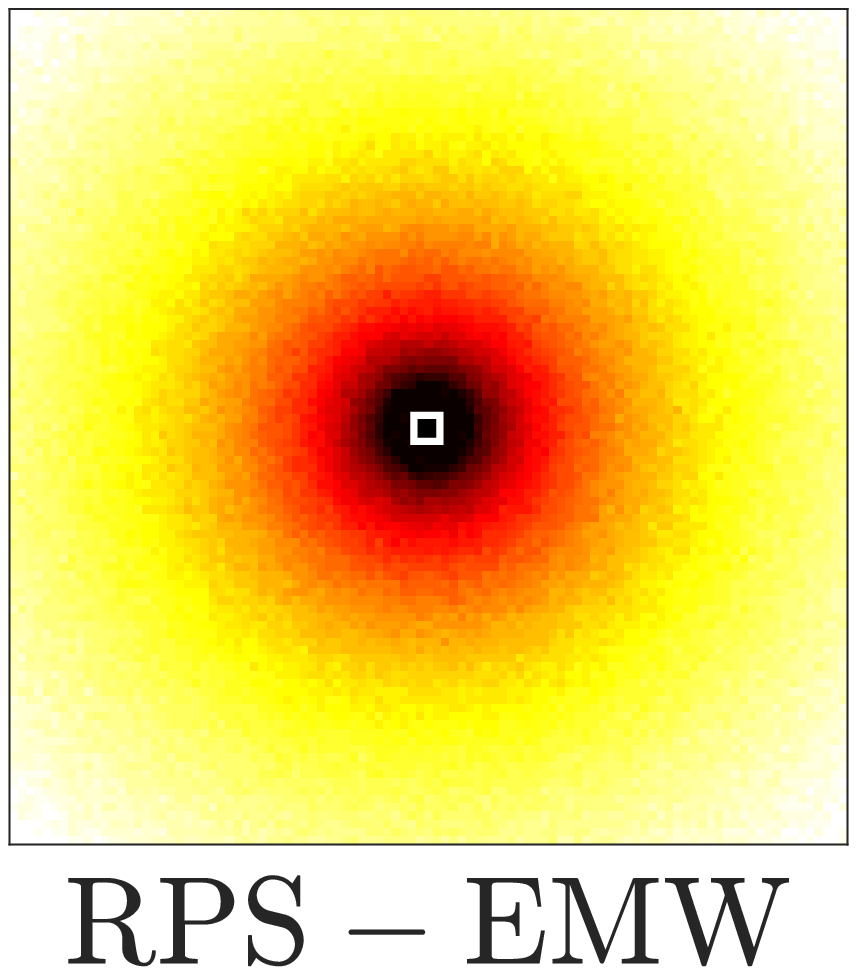}\ \ \!\!\!\! \ 
		\includegraphics[width=0.085\textwidth]{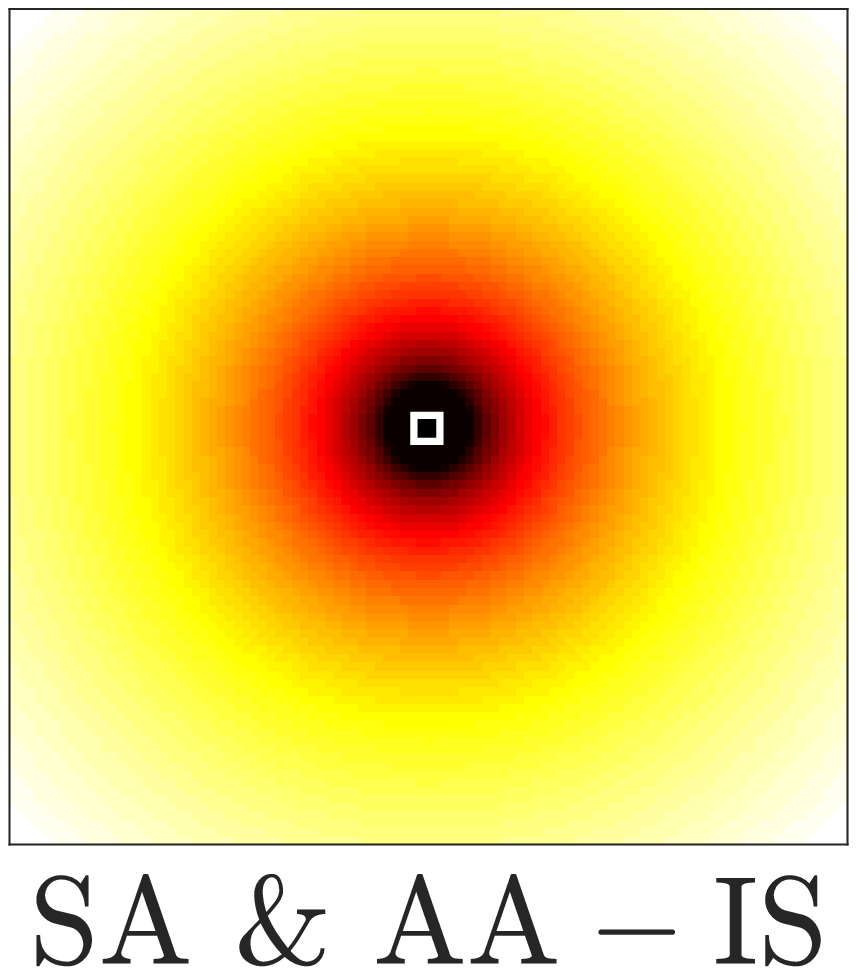}\ \ \!\!\!\! \ 
		\includegraphics[width=0.085\textwidth]{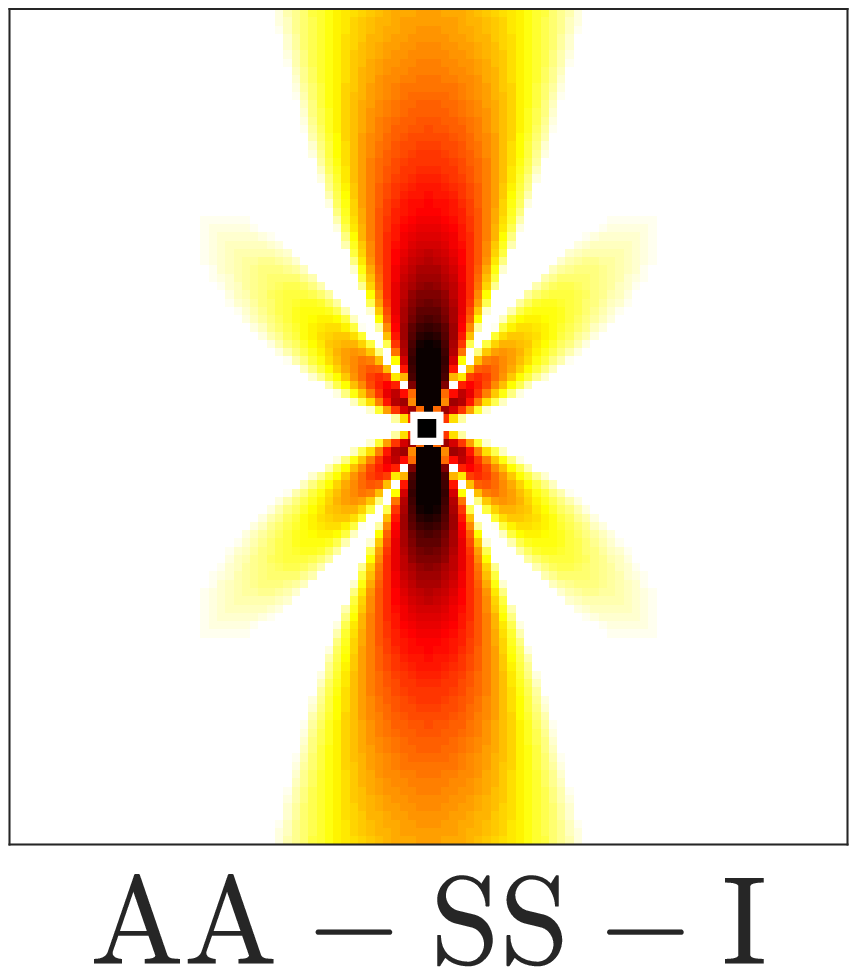}\ \ \!\!\!\! \ 	
		\includegraphics[width=0.085\textwidth]{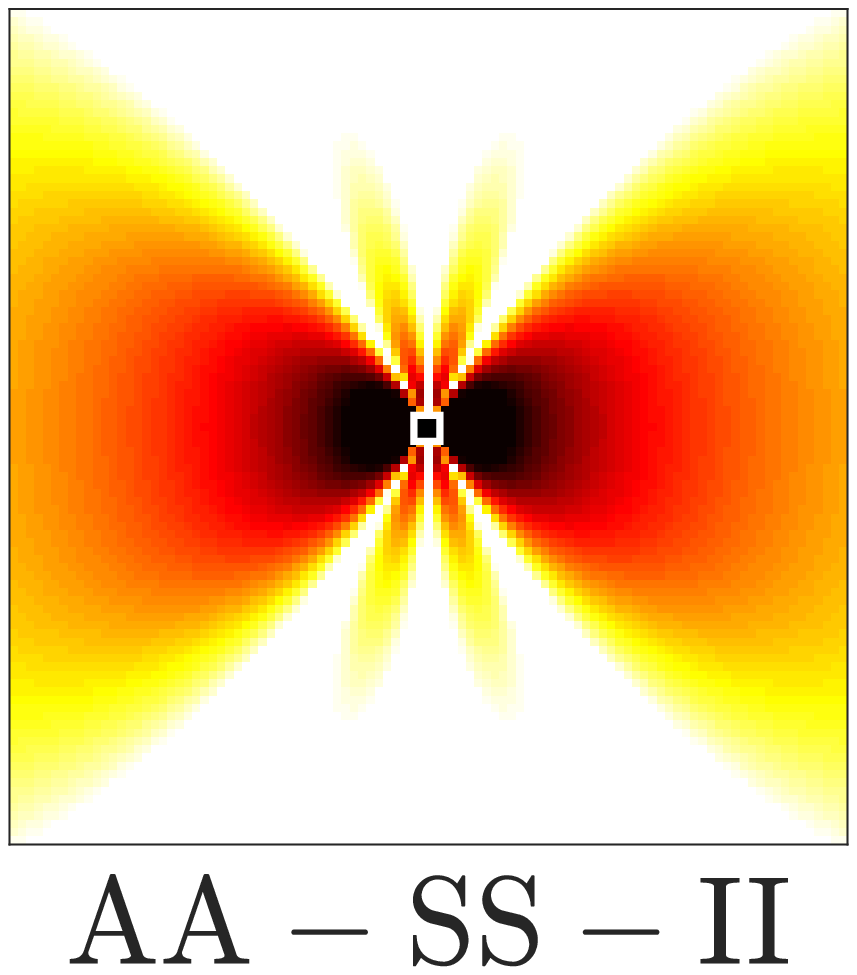}\ \ \!\!\!\! \ 
		\includegraphics[width=0.085\textwidth]{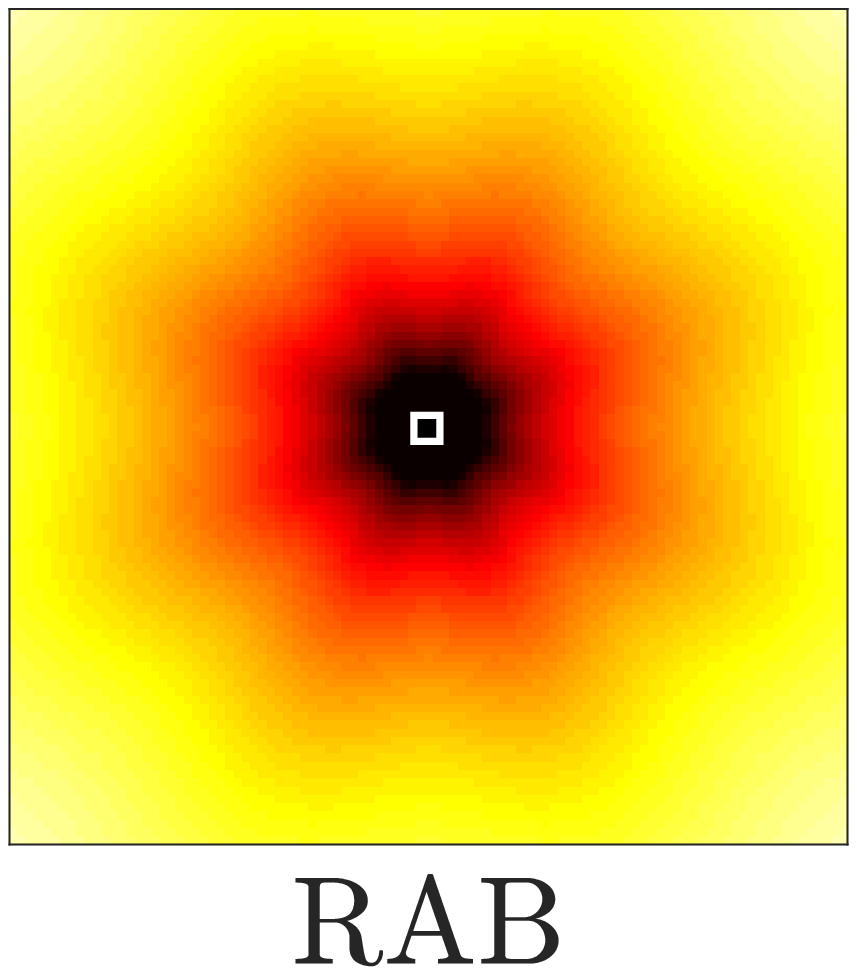}		
		\caption{Heatmap of the average RF energy availability in dBm under the discussed CSI-free WET schemes.}
		\label{Fig6}
	\end{figure}
\begin{figure}[t!]
	\centering
	\includegraphics[width=0.48\textwidth]{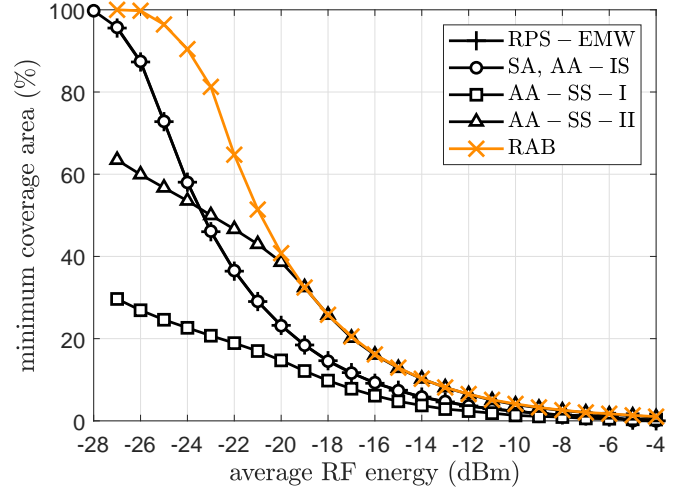}
	\caption{Area coverage for different  average RF energy requirements. The curves corresponding to RPS-EMW and SA, AA-IS are overlapping.}
	\label{Fig7}
\end{figure}
    \subsection{On the Area Coverage}\label{coverage}
    First, Figs.~\ref{Fig6} and \ref{Fig7} show the system performance in terms of average RF power availability in a $20 \times 20\ \mathrm{m^2}$ area served by a PB at the area center that uses the different state-of-the-art CSI-free WET schemes overviewed in Section~\ref{csiArt} and our proposed RAB strategy (Section~\ref{bar}). 
    Specifically, Fig.~\ref{Fig6} illustrates a heatmap of the area energy availability, and evidences the omnidirectional radiation pattern attained under RPS-EMW, SA and AA–IS, while AA-SS schemes favor certain spatial directions (ULA boresight, and $90^\circ$ offset from the boresight, directions in case of AA-SS-I and AA-SS-II, respectively), and the proposed RAB reaches a quasi-omnidirectional pattern after averaging over all ULA rotations. Meanwhile, a quantitative information on the average RF energy availability is shown in Fig.~\ref{Fig7}. For instance, observe that while AA–SS-I allows covering up to $20\%$ of the area with a guarantee of $-22$ dBm ($\sim 6\ \mu$W), the coverage increases to $37\%$ when using SA, AA-IS and RPS-EMW, and it might even reach $47\%$ and $65\%$ under the AA-SS-II and RAB schemes, respectively. Moreover, as energy requirements loosen, the area can be more easily served by schemes providing omnidirectional or quasi-omnidirectional radiation patterns as in case of RPS-EMW, SA, AA-IS, and RAB. However, the latter is shown to be the clear winner in every situation, for either loose or tight energy requirements. The disadvantage of RAB with respect to state-of-the-art CSI-free schemes in \cite{Clerckx.2018,LopezAlves.2019,LopezMontejo.2020,LopezMonteiro.2021} is that it requires an extra hardware component, i.e., a rotor, which may significantly raise the hardware cost and energy consumption. On the bright side, by properly tuning the PB's transmit power, the EMF exposure in the charging area can be kept at bay  for a broader range of EH requirements at the remote nodes.    
    \subsection{On the Impact of $|\mathcal{S}|$ and $M$}\label{SM}
    From now on, we discard RPS-EMW and AA-SS-I schemes since the performance of the former is extremely similar to that of SA and AA-IS, while the latter performs extremely poor in general deployments. Additionally, we consider a finite set of EH devices randomly and uniformly distributed in a 10 m-radius area around the PB, which allows also to compare our proposal not only to the remaining state-of-the-art CSI-free WET schemes, but also to a full-CSI based strategy without antenna rotation.\footnote{Note that a properly designed full-CSI precoder  additionally exploiting antenna rotation would, in principle, outperform  all these schemes (at least if devices' energy expenditures is not taken into account). However, such a design is not straightforward and it has been left for future work as it definitely constitutes an interesting and promising research direction.} Such latter scheme consists of a precoder designed to form beams to reach the EH devices with maximum fairness, i.e., no device is expected to benefit more from PB's WET than others. This problem can be stated as a semi-definite program (SDP) \cite{LopezMonteiro.2021}, which in turn can be solved efficiently. For the analysis performance of this CSI-based strategy, we are not considering the power consumed in the CSI acquisition phase, which would further tilt the scale in favor of the CSI-free schemes. 
	
	Fig.~\ref{Fig8} depicts the average worst-case  RF energy available at a varying number of EH devices. Observe that the performance of all schemes worsens  as the number of devices increases since i) the chances of having devices farther from the PB increase, and ii) the energy beams (in case of AA-IS-II and full-CSI) become less capable of efficiently reaching the devices. From Fig.~\ref{Fig8}, we can make the following specific observations:	
	\begin{figure}[t!]
		\centering
		\includegraphics[width=0.48\textwidth]{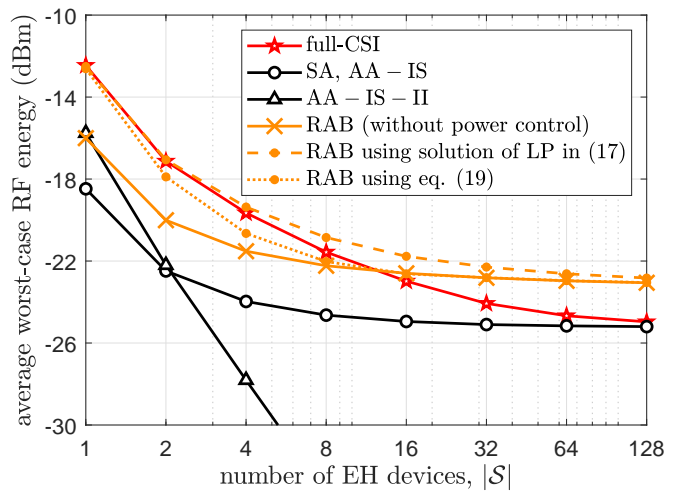}
		\caption{Average worst case RF energy at the set $|\mathcal{S}|$ of IoT devices. We set $M=4$.}
		\label{Fig8}
	\end{figure}
	\begin{itemize}
		\item the proposed RAB scheme outperforms all the state-of-the-art CSI-free WET schemes. Even when serving a single EH device, RAB is preferable to AA-SS-II since the device's performance under the latter may be seriously affected/boosted depending on the device's position; 
	\item the performance gap between the un-optimized RAB and SA (and AA-IS) matches $10\log_{10}(0.85\sqrt{4})=2.3$ dB, which corresponds to the EH gains over single antenna transmissions claimed in \eqref{avP}. This is because SA and AA-IS do not provide average EH gains, but only diversity gains;
	\item even without further optimization, RAB eventually outperforms the full-CSI (without antenna rotation) based precoder when serving a sufficiently large number of devices, i.e., larger compared to the number of antenna elements. For the adopted system configuration, the scenario with 12 EH devices performs already better with the un-optimized RAB than with the full-CSI precoder;
	\item optimized RAB provides additional performance gains, specially when serving a relative small number of devices. This is because the devices' spatial deployment is less homogeneous when serving a limited set of EH devices, thus, a suitable per-rotation(beam) power control can favor the directions where devices are farther from the PB and unleash important gains;
	\item RAB optimized using the solution of $\mathbf{P}$ in \eqref{P} outperforms all its competitors, including the full-CSI precoder, in all the cases. Important gains over the un-optimized RAB can even be obtained when using the sub-optimum but simple power control specified in \eqref{popt};
	\item performance gains coming from exploiting devices' positionining information vanish as the number of devices grows unbounded, i.e., the performance under the optimized RAB matches that of the un-optimized RAB as $|\mathcal{S}|\rightarrow\infty$.
	\end{itemize}

\begin{figure}[t!]
	\centering
	\includegraphics[width=0.48\textwidth]{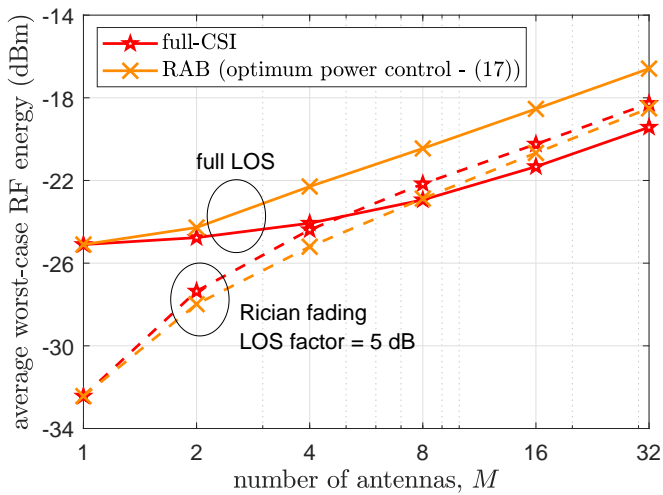}
	\caption{Average worst case RF energy at a set of 32 EH devices as a function of $M$.}
	\label{Fig9}
\end{figure}

As shown in Fig.~\ref{Fig8}, the optimized RAB scheme, using the power control that comes from solving $\mathbf{P}$ \eqref{P}, easily outperforms the full-CSI without antenna rotation design. However, this claim holds strictly only when assuming full LOS channels, and can be also appreciated in Fig.~\ref{Fig9} for a varying number of transmit antenna elements at the PB powering 32 EH devices. However, the situation may be somewhat different when considering channels under the influence of some NLOS signal components. To illustrate this, Fig.~\ref{Fig9} includes performance curves corresponding to a scenario with channels subject to Rician fading with LOS factor of 5 dB, which may be rather pessimistic for typical short range WET. Observe now that under such channel conditions, the full-CSI precoder can slightly outperform (without considering the training energy costs) our proposed optimized CSI-free RAB solution. Obviously, the performance gap increases in favor of the full-CSI precoder as the LOS factor decreases, thus, a CSI-based beamformer  may be strictly required for serving the EH devices under such harsh channel conditions.

\subsection{On the SAR-constrained Performance}
Let us return to the LOS scenario and consider now that beams corresponding to angular rotations $j=1,2$ need to be limited in power due to the presence of a human. We assume the PB is equipped with $M=4$ antenna elements, and a single SAR matrix is provided (for simplicity and to ease our exposition, thus we avoid index $z$), which is given by \cite{Zhang.2020}
	\begin{align}
	\mathbf{S}=\begin{bmatrix}
	1.6 & -1.2\mathbbm{i} & -0.42 & 0\\
	1.2\mathbbm{i} & 1.6 & -1.2\mathbbm{i} & -0.42\\
	-0.42 & 1.2\mathbbm{i} & 1.6 & -1.2\mathbbm{i}\\
	0 & -0.42 & 1.2\mathbbm{i} & 1.6
	\end{bmatrix}.
	\end{align}
	By using \eqref{SAR1}, we have that $\text{SAR}_j=1.18p_j$, thus $p_j\le \delta/1.18,\ j=1,2$, which can be directly incorporated to $\mathbf{P}$ in \eqref{P} (and even to \eqref{popt} with some minor changes). In case of the traditional full-CSI precoding implementation adopted here for benchmarking, the generated energy beams are not location-specific, specially for $M\ll |S|$, thus, the SAR constraint needs to be applied over the entire set of beams. This is,
	\begin{align}
	\text{SAR}_{\text{full-csi}}=\sum_{k=1}^{K}\mathbf{v}_k^H\mathbf{S}\mathbf{v}_k=\Tr(\mathbf{S}\mathbf{V}_{\text{full-csi}}),
	\end{align}
	where $\mathbf{V}_{\text{full-csi}}=\sum_{k=1}^{K}\mathbf{v}_k\mathbf{v}_k^H$ and $\Tr(\mathbf{V}_\text{full-csi})=p_T$. Then, $\Tr(\mathbf{S}\mathbf{V}_{\text{full-csi}})\le \delta$ can be efficiently included into the optimum full-CSI SDP precoding optimization specified in \cite{LopezMonteiro.2021} to find $\mathbf{v}_k^\text{opt}, \forall k$.
	
\begin{figure}[t!]
	 	\centering
	 	\includegraphics[width=0.48\textwidth]{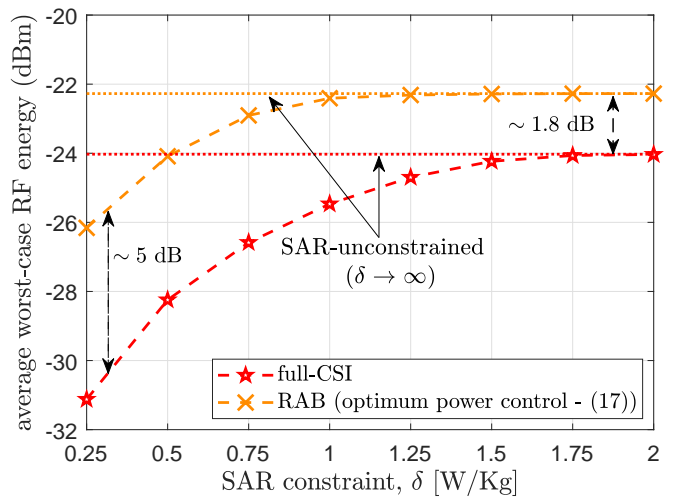}
	 	\caption{Average worst case RF energy at a set of 32 EH devices as a function of SAR constraint. We set $M=4$.}
	 	\label{Fig10}
	 \end{figure}

	Results in Fig.~\ref{Fig10} evidence that the proposed optimized RAB scheme outperforms more easily the full-CSI scheme as SAR constraints become more stringent, e.g., with performance gaps from $\sim 1.8$ dB to $\sim 5$ dB when SAR limits decrease from infinity (no SAR constraints) to $0.25$ [W/Kg]. 	 
	Herein, we would like to make notice that common SAR limits usually vary within the interval $1 \le \delta\le 2$ [W/kg] \cite{Zhang.2020, Chan.2001,British.1999,Iphone}. For such an interval, the performance under the full-CSI scheme is seriously affected, while RAB is not longer SAR-limited  for $\delta\ge 1.25$ [W/kg] under the adopted system configuration.
	\section{Conclusion}\label{conclusions}
	In this article, we proposed a novel CSI-free WET scheme, referred to as RAB, to be adopted by a multi-antenna PB to wirelessly power a massive set of EH IoT devices. RAB uses a properly designed CSI-free beamformer combined with a continuous and smooth antenna array rotation to provide, to all devices, an average EH gain (compared to state-of-the-art omnidirectional CSI-free WET schemes) proportional to the square-root of the number of PB's antenna elements.  We showed that even when a smooth rotation is not possible, the PB may be equipped with a stepper rotor with a number of equally-spaced steps matching the number of array antennas. A rotation-specific power control mechanism was proposed to i) fairly optimize the WET process if devices' positioning information is available, and ii) make RAB to be SAR-compliant, thus, avoiding risks to human health. It was shown that the optimum power control comes from solving an LP, e.g., using standard optimization techniques and tools, while a sub-optimal closed-form power allocation, which allows simpler implementation, was also derived. We discussed important practicalities related to RAB, e.g., its robustness against NLOS compared to other CSI-free WET schemes, and its generalizability to scenarios where the PB uses other than a ULA topology. 
	We showed that RAB outperforms all the state-of-the-art CSI-free WET schemes available in the literature; and even without optimized power control its performance may be superior to that provided by a full-CSI precoder when the number of devices grows larger than the number of PB's antenna elements. Meanwhile, by adopting the proposed rotation-specific power control, RAB can more easily outperform the full-CSI precoder. Finally, we illustrated that RAB is less sensitive to SAR constraints than the full-CSI based design.
		
An interesting research direction for future work is to investigate the performance of RAB under non-linear (more practical) EH models and for different array topologies. Note that important EH gains are also expected for other antenna array topologies, and the optimum array implementation needs to be defined. Additionally, since rotation provides another degree of freedom for performance improvements, it would be interesting to design an alternative full/limited CSI-based RAB scheme, which may provide important performance gains over traditional (static/fixed PBs) CSI-based precoders.

	\bibliographystyle{IEEEtran}
	\bibliography{IEEEabrv,references}
\end{document}